\documentclass[aps,onecolumn,showpacs,amsmath,amssymb,pra,reprint]{revtex4-1}  
\usepackage{graphicx}%
\usepackage{dcolumn}
\usepackage{bm}
\usepackage{braket} 
\usepackage{color} 
\begin{document} 
\title{ Femtosecond time-resolved dynamical Franz-Keldysh effect} 
\author{T. Otobe$^{1}$, Y. Shinohara$^{2}$, S. A. Sato$^3$, and K. Yabana$^{1,3,4}$}
\affiliation{$^1$Kansai Photon Science Institute, Japan Atomic Energy Agency, Kizugawa, Kyoto 619-0615, Japan\\ 
$^2$Max-Planck Institut f\"{u}r Mikrostrukturphysik, Weinberg 2, D-06120, Halle, Germany\\ 
$^3$Graduate School of Pure and Applied Sciences, University of Tsukuba, Tsukuba  305-8571,Japan\\ 
$^4$Center for Computational Sciences, University of Tsukuba, Tsukuba 305-8577,
Japan}

\begin{abstract} 
We theoretically investigate the dynamical Franz-Keldysh effect in 
femtosecond time resolution, that is, the time-dependent modulation 
of a dielectric function at around the band gap under an irradiation 
of an intense laser field. We develop a pump-probe formalism 
in two distinct approaches: first-principles simulation based on 
real-time time-dependent density functional theory and analytic 
consideration of a simple two-band model. We find that, 
while time-average modulation can be reasonably described by  
the static Franz-Keldysh theory, a remarkable phase shift is found 
to appear between the dielectric response and the applied electric field. 
\end{abstract} 
\maketitle 
\section{introduction} 
In last two decades, intense coherent light of different 
characteristics has become available owing to advances
in laser sciences and technologies. Ultrashort  
laser pulses can be as short as a few tens of attosecond,  
forming a new field of attosecond science \cite{atto01}.  
Intense laser pulses of mid-infrared (MIR) or THz frequencies  
have also become available recently \cite{HDBT11, Chin01}.  
Employing these extreme sources of coherent light, it is  
possible to investigate the optical response of materials in real 
time with a resolution much less than an optical  
cycle\cite{atto01,Hirori11,Krausz13,Schultze13,Schultze14,Novelli13}.  
 
The dielectric function $\varepsilon(\omega)$ is the most fundamental  
quantity characterizing the optical properties of matter.  
Modulation of the dielectric function  $\varepsilon(\omega)$  in the presence of  
electromagnetic fields has been a subject of investigation  
for many years. The change under a static electric field is known  
as the Franz-Keldysh effect (FKE)   
\cite{Franz58,Keldysh58,Tharmalingam63,Seraphin65,Nahory68,
Shen95,Sipe10,Sipe15}, 
and that under an alternating electric field is known as  
the dynamical FKE (DFKE)  
\cite{Yacoby68, Jauho96, Nordstorm98, Ajit04,Mizumoto06, Shambhu11}. 
An important parameter which distinguishes DFKE from 
the static FKE is the adiabaticity parameter $\gamma = U_p/\Omega$,  
where $U_p= e^2 E^2/4 \mu \Omega^2$ is the  
ponderomotive energy, and $\Omega$ is the frequency 
of the field, $\mu$ is the reduced mass of the electron, 
and $E$ is the electic field  \cite{Nordstorm98}. 
A multi-photon picture applies for $\gamma << 1$, and 
a static FKE picture is appropriate for $\gamma >> 1$. 
Laser pulses having $\gamma \sim 1$ is an intriguing regime 
where novel and unobvious DFKE phenomena are expected. 
 
In previous investigations of DFKE, the main focus was on 
the modulation of the optical response averaged over times much longer  
than the optical cycle, examining the time-averaged fine structure \cite{Jauho96}  
and  shifts in excitation structures \cite{Nordstorm98}.
In the present paper, we examine DFKE in time domain, with a  
resolution much less than the cycle of the applied optical field.  
The DFKE response in subfemtosecond time resolution is very relevant
to ultrafast optical switching in the teraherz or even 
petahertz ($10^{15}$ hertz) domain \cite{Novelli13,Krausz13}. 
A first experimental report on the DFKE with a femtosecond time resolution  
has recently been given by Novelli \textit{et al.} \cite{Novelli13} for GaAs, 
employing an intense pump pulse of THz frequency.  
They observed an interesting time shift between  
the pump pulse and the modulation of dielectric function,  
but the mechanism of the observed time profile was not understood.
To uncover the physics of time-resolved DFKE, we develop  
a pump-probe formalism in two different theoretical approaches:  
first-principles numerical simulations based on time-dependent 
density functional theory (TDDFT \cite{Runge84} ) and analytic investigation 
for a two-band model. Combining two approaches,  
we can understand not only the strength of the modulation but
the phase with respect to the pump field as well.

 The organization of the present article is as follows.
 In section II, we present formalism and results of our first-principles 
 calculations for the time-resolved DFKE.
 In section III, we develop an analytical formalism for the time-resolved 
 DFKE employing a parabolic two-band model.
 In section IV, a summary will be presented.
 \section{First-principles pump-probe calculation}
 \subsection{Formalism}
 \label{sec:Formalism}
In real-time TDDFT, we describe electron dynamics in a unit  
cell of a crystalline solid under a spatially-uniform electric field $\vec E(t)$. 
The method has been applied for calculations of linear optical 
responses \cite{Bertsch00} and nonlinear electronic excitations  
by intense laser pulses \cite{Otobe08, Otobe09, Shino10-2, Shino12, Otobe12, 
Sato14, Wachter14}.  
Treating the field by a vector potential 
\begin{equation}
\vec A(t)=-c\int^t dt' \vec E(t'), 
\end{equation}
the electron dynamics in the unit cell of solid is described by the following 
time-dependent Kohn-Sham (TDKS) equation \cite{Bertsch00} : 
\begin{eqnarray} 
 &&i \frac{\partial}{\partial t} \psi_i(\vec{r},t)=
\left[ \frac{1}{2m_e} \left( \vec p + \frac{e}{c} \vec A(t) \right)^2 \right. \nonumber \\
&& \left.
+ V_{ion}(\vec r) + V_H(\vec r,t) + V_{xc}(\vec r,t) \right] \psi_i (\vec r,t), 
\label{TDKS} 
\end{eqnarray} 
where $m_e$ is the electron mass, $V_{ion}$ is the electron-ion potential for which
we use  a norm-conserving pseudpotential \cite{TM91,Kleinman82}, and 
$V_H(\vec r,t)$ and $V_{xc}(\vec r,t)$ are electron-electron Hartree and 
exchange-correlation potentials, respectively.  For the exchange-correlation potential, 
we employ an adiabatic  local density approximation, using the same functional form of 
the potential for both ground state and time evolution calculations \cite{PZ81}.
Since the Kohn-Sham Hamiltonian in Eq. (\ref{TDKS}) has the lattice periodicity at
each time, we may introduce time-dependent Bloch wave function,
$\psi_i(\vec r,t) = e^{i\vec k \cdot \vec r} u_{n\vec k}(\vec r,t)$. In practice,
we calculate the time evolution of the Bloch wave functions.

We calculate electron dynamics in diamond, using a cubic unit cell  
containing eight carbon atoms. 
The TDKS equation is solved in  
real time and real space. 
The real-space grids of $22^3$ is used  
for the unit cell, and $32^3$ grids for the $k$-points.  
The Taylor expansion method is used for the time evolution \cite{Yabana96}  
with a time step of $\Delta t = 0.02$ in atomic unit.  
The number of time steps is typically 70,000. 
An important output of the calculation is the
average electric current density as a function of time, $\vec J(t)$.
It is given by 
\begin{equation} 
\vec J(t) = -\frac{e}{m_eV}  \int_{V} d\vec r \sum_i 
{\rm Re} \psi_i^* \left( \vec p + \frac{e}{c}\vec A(t) \right) \psi_i + J_{NL}(t), 
 \label{current} 
\end{equation} 
where $V$ is a volume of the unit cell.  
$\vec J_{NL}(t)$ is the current caused by non-locality of the pseudopotential. 
 
It should be mentioned that $\vec A(t)$  in the TDKS equation (\ref{TDKS})  
is the vector potential in the medium and not that of the incident
pulse in the vacuum. 
As we discussed in Ref.\cite{yabana12}, the relation
between the two depends on the macroscopic 
shape of the materials as well as the direction of the polarization.
Exchange-correlation effects may also appear in the vector potential $\vec A(t)$ 
in time-dependent current density functional theory \cite{Vignale96}, 
which we ignore for simplicity. 

To examine the DFKE, we carry out simulations solving the TDKS 
equation (\ref{TDKS}) 
 including both pump and probe electric fields 
in the vector potential $\vec A(t)$ \cite{Sato14}. 
We assume that both pump and probe electric fields are linearly 
polarized and are orthogonal to each other. 
We denote the pump electric field as $E_P(t)$ and the probe 
electric field as $E_p(t)$. 
The probe electric field is assumed to be weak enough to be treated  
by the linear response theory. We denote the electric current caused by  
the probe field as $J_p(t)$, which is assumed to be parallel to the
direction of the probe electric field. They are related by the time-domain  
conductivity $\sigma(t,t')$ as 
\begin{equation} 
J_p(t) = \int_{-\infty}^t dt' \sigma(t,t') E_p(t'). 
\label{def_sigma} 
\end{equation} 
We note that the conductivity $\sigma(t,t')$ depends on both 
times $t$ and $t'$ rather than just time difference
$t-t'$ due to the presence of the pump pulse.
 
We derive a frequency-dependent conductivity at time $T$ 
from the conductivity $\sigma(t,t')$. 
We first note that, in the absence of the pump electric field,  
the frequency-dependent conductivity  
$\tilde{\sigma}(\omega)$ is related to the Fourier transforms  
of the electric field and the induced current as
$\tilde{\sigma}(\omega) =  
\int dt e^{i\omega t} J_p(t)/\int dt e^{i\omega t} E_p(t)$,
where we may use any time profile for the probe electric field, $E_p(t)$.
We employ this relation in the presence of the strong pump field. 
To introduce the time $T_p$ to explore the response,  a pulsed  
electric field is employed for $E_p(t)$ whose envelope shows  
a maximum at $t=T_p$.
Namely, we use the following relation to define the time-resolved,
frequency-dependent conductivity $\sigma(T_p,\omega)$,
\begin{equation}
\tilde{\sigma}(T_p,\omega) =  
\frac{\int dt e^{i\omega t} J_p(t)}{\int dt e^{i\omega t} E_p(t)}.
\label{TD_conductivity}
\end{equation}
Thus, in this consideration, the probe electric field has a dual role,  
to specify the time $T_p$ at which we explore the response of the  
medium and to distort the medium to examine responses of  
the system. We consider that this dual role of the probe pulse 
will be adopted to measure the transient dielectric functions 
in most experiments. 
 
\subsection{Numerical results}
 
In the calculations below, we use the following electric fields. 
The pump field is of the form
$E_P(t) = E_{0,P} f_P(t) \sin \Omega t $ with the central
angular frequency $\Omega$ set to $\Omega=0.4$ eV; its
direction is along the [001] axis.
The field is turned on adiabatically described by the function $f_P(t)$,
\begin{eqnarray} 
\label{pump-laser} 
f_{P}(t) &=& \nonumber \left\{ \begin{array}{ll} 
    \sin^2\left( \frac{\pi }{2 T_P}t\right)  
& (0<t<{T_P}) \\ 
    1 & (t \ge T_P) ,
  \end{array} \right. \nonumber \\ 
\end{eqnarray}
where $T_P$ is set to 10 fs.  
The probe field is of the form
\begin{equation}
E_{p}(t) = E_{0,p}  \sin(\omega_p t) 
\exp\left(-\frac{(t-T_p)^2}{\eta^2}\right),
\end{equation} 
oriented in the [100]
direction.  The average frequency $\omega_p$ is set to 5.6 eV, 
which is equal to the calculated band gap energy  
of diamond in LDA. The field strength is set to 
$E_{0,p}=2.7 \times 10^{-3}$ MV/cm, which is small enough
to probe the linear response of the medium.  
The pulse
duration $\eta$ is set to $\eta= 0.7$ fs.  With such a
short duration, 
we may scan the spectrum of broad frequency region around  the  
band gap from a single pump-probe calculation.  

The frequency-dependent conductivity is calculated using
Eq. (\ref{TD_conductivity}) with a slight modification,
\begin{equation} 
\tilde{\sigma}(T_p, \omega) =  
\frac{\int dt e^{i\omega t} g(t-T_p) J_p(t)}{\int dt e^{i\omega t} E_p(t)}, 
\label{conductivity} 
\end{equation} 
where we introduced a filter function, $g(t)$, \cite{yabana06} to suppress spurious 
oscillations which arise from a finite time period of the simulation. 

\begin{figure} 
\includegraphics[width=90mm]{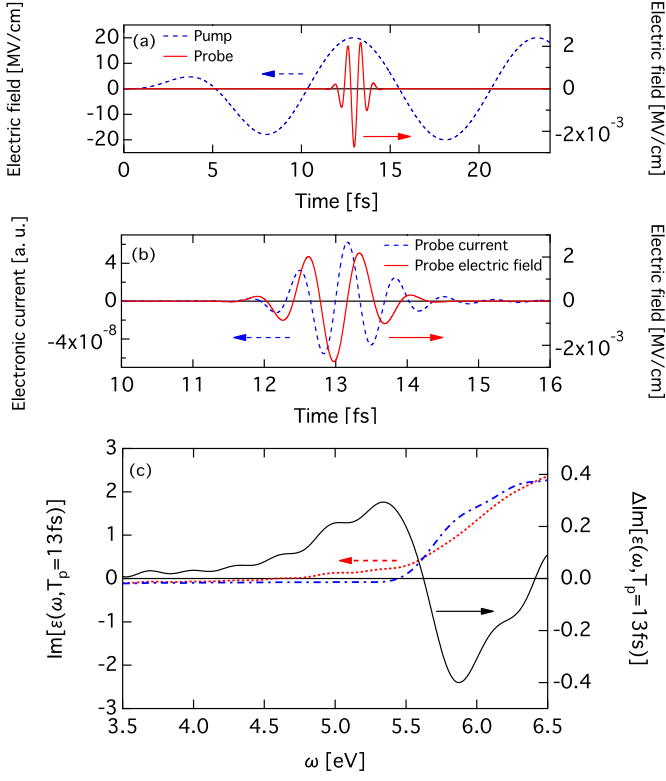} 
\caption{\label{fig:Field_Cur} (a) Pump (blue-dashed line) and  
probe (red-solid line) electric fields are shown. 
(b) The electronic current (blue-dashed line)  
induced by the probe electric field (red-solid line). 
(c) The imaginary part of the dielectric function in the 
presence of the pump field,  
${\rm Im}[\varepsilon(\omega, T_p=13fs)]$ (red-dashed line), 
and in the absence of the pump field,  
${\rm Im}[\varepsilon(\omega)]$ (blue-dash-dot line).  
Black-solid line shows the difference, 
${\rm Im}[\varepsilon(\omega, T_p=13fs)]- 
{\rm Im}[\varepsilon(\omega)]$. 
 } 
\end{figure}

Figure~\ref{fig:Field_Cur} shows an example of our pump-probe calculations. 
In (a), the electric fields of the pump field $E_P(t)$ and the probe pulse $E_p(t)$ 
are presented: The blue-dashed line shows the electric fields of the 
pump field $E_P(t)$, and the red-solid line shows the electric field of the  
probe pulse $E_p(t)$. 
The magnitude of the pump electric field , $E_{0,P}$,   
is set to 20 MV/cm. The probe pulse is applied at a time when the  
magnitude of the pump field is maximum.
 The electric current induced by the  
probe pulse, $J_p(t)$, is shown by the blue-dashed line in (b). 
The conductivity is calculated from the probe current using  
Eq. (\ref{conductivity}), and then converted to the dielectric function 
using a formula 
\begin{equation}
\epsilon(\omega)=1+4\pi i \frac{\sigma(\omega)}{\omega},
\end{equation}
 which is valid in the 
presence of the strong pump field. 
The imaginary part of the dielectric function,  
Im[$\varepsilon(\omega, T_p=13fs)$], is shown in  (c). 
The red-dotted and blue dot-dashed lines present the dielectric function with 
and without the pump field, respectively. 
A change of the imaginary part of the dielectric function, 
$\Delta$ Im$[\varepsilon(\Omega, T_p=13fs)]=$Im$[\varepsilon(\Omega, T_p=13fs)]-{\rm Im}[\varepsilon(\omega)]$, 
is shown  by the black-solid line. It indicates modulations of an exponential tail below and 
an oscillation above the band gap ($\omega=$ 5.5 eV).

\begin{figure} 
\includegraphics[width=90mm]{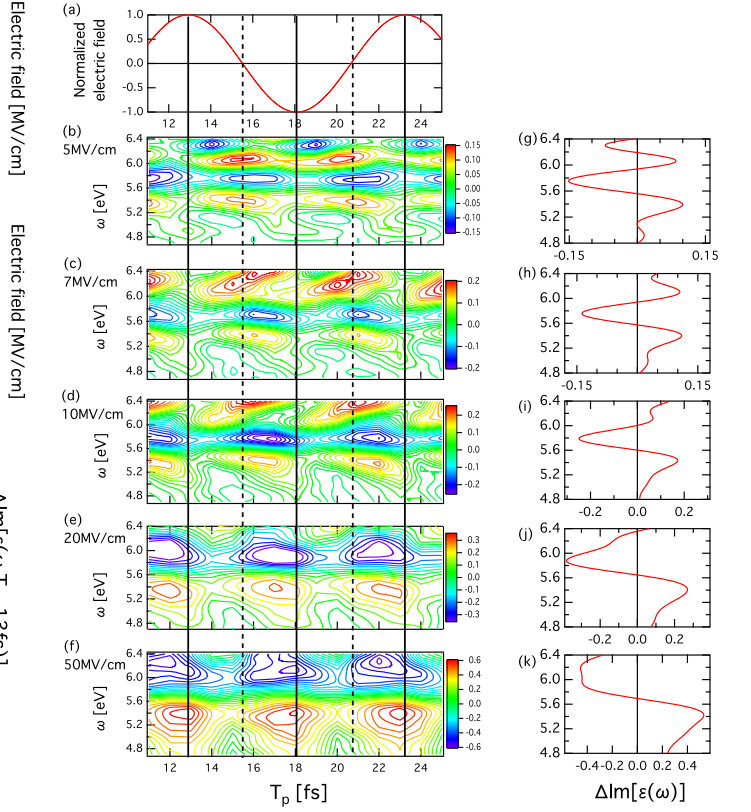} 
\caption{\label{fig:TD_Sigma} Contour plots (left) and their time  
averages (right) of $\Delta{\rm Im}[\varepsilon(\omega, T_p)]$ under  
the MIR pump field of the intensity of (b,g) 5, (c,h) 7, (d,i) 10, (e,j) 20,  
and (f,k) 50 MV/cm. The time dependence of the pump electric field is  
shown in (a).  In panels (a)-(f), the vertical solid (dashed) lines indicate  
the time of maximum (zero) of the pump electric field.} 
\end{figure} 
 

Figure \ref{fig:TD_Sigma} shows changes of the imaginary part of the 
dielectric function caused by the strong pump field,  
$\Delta{\rm Im}[\varepsilon(\omega, T_p)] 
={\rm Im}[\varepsilon(\omega, T_p)]-{\rm Im}[\varepsilon(\omega)]$. 
In (a), time profile of the pump electric field is shown. 
In (b) - (f), changes of imaginary part of the dielectric function 
are shown in contour plots for four intensities. 
Horizontal axis is the time $T_p$ and the vertical axis is the frequency $\omega$.  
In (g)-(k), modulations averaged over time are 
shown as a function of frequency.

We first look at the case of strongest pump electric field, $E_{0,P}=50$ MV/cm 
,shown in (f) and (k). 
We find that an increase of the absorption below and a decrease above the
band gap are seen when the magnitude of the pump electric field is
close to the maximum. The modulation is small when the electric field
is close to zero. This fact indicates the static FKE appears instantaneously
following the change of the pump field. We confirmed the modulation is
well fitted by the static FKE formula if we assume the effective mass of
$\mu=0.25 m$. With this value of the effective mass, the adiabatic
parameter $\gamma$ is 29.5, much larger than unity, which is consistent
with the appearance of the static FKE.

As the field intensity decreases, we find changes in two aspects,  
one along the frequency  axis and the other along the time axis. 
Above the band gap, we find an oscillatory behavior in the 
frequency  direction. This is clearly seen in the time-average 
behaviors shown in (g)-(k). For example, at the electric field  
of 10 MV/cm,
 shown in (d) and (i), 
the modulation is negative between 
5.6 eV (band gap) to 6.0 eV. It then becomes positive above 
6.0 eV.  
 
Along the time  axis, we can see a striking change in  
the phase between the modulation and the pump electric field. 
At the strongest electric field (f), the modulation and the pump 
electric field is in phase, as mentioned above. As the intensity of the 
pump field decreases, a phase shift forward in time is seen. 
The amount of the phase difference increases as the magnitude of the pump  
field decreases, as seen from (b) to (f). 
At the lowest intensity (b), the modulation signals appear at times when 
the pump field is close to zero. 
The parameter $\gamma$ for these intensities is (b) 0.29, (c) 0.58,  
(d) 1.19, and (e) 4.76. Thus the phase change becomes appreciable  
when the $\gamma$ value is around and below unity.

 
\section{Analytic consideration}

\subsection{General formula for the conductivity in the presence
of a strong field}

We next develop an analytic consideration to understand 
behaviors of the DFKE signals seen in the  numerical simulation.
In the following developments, we consider a simplified description:
electron dynamics in the presence of pump and probe fields is 
assumed to be described by a time-dependent Schr\"odinger 
equation for a single electron,  
\begin{equation}
 i\frac{\partial}{\partial t} \psi_i(\vec r,t) 
= \left[\frac{1}{2m_e}\left(\vec p+\frac{e}{c}\vec A(t)\right)^2+V(\vec r) \right] \psi_i(\vec r,t) 
\label{TDSE}
\end{equation}
where $V(\vec r)$ is a time-independent, lattice periodic potential.
We thus ignore the time-dependence of the Kohn-Sham Hamiltonian
of Eq. (\ref{TDKS}), except for the vector potential.
We express the solution of this equation using time-dependent 
Bloch function $v_{n\vec k}(\vec r,t)$ as
$\psi_i(\vec r,t)=e^{i\vec k \vec r} v_{n \vec k}(\vec r,t)$.  

We further assume that, in the presence of the pump field
described by a vector potential $\vec A_P(t)$,  the solution of 
Eq. (\ref{TDSE}) is well approximated by the so-called Houston 
function \cite{Yacoby68,Houston}. 
Using static Bloch orbitals  $u_{n\vec k}(\vec r)$ and orbital energies 
$\epsilon_{n\vec k}$ which satisfy
\begin{equation}
\left[\frac{1}{2m_e}\left(\vec p+\vec k \right)^2+V(\vec r) \right] u_{n\vec k}(\vec r)
= \epsilon_{n\vec k} u_{n\vec k}(\vec r),
\end{equation}
the Houston function is expressed as 
\begin{equation} 
 w_{n\vec{k}}(\vec{r},t)=u_{n \, \vec k_P(t)}(\vec{r}) 
 \exp \left[-i \int^t \epsilon_{n \,\vec k_P(t')} dt' \right],
 \end{equation} 
where $\vec k_P(t)$ is defined by $\vec k_P(t) = \vec k + e\vec A_P(t)/c$.

We then consider the solution of Eq. (\ref{TDSE}) in the presence of 
both pump and probe fields, $\vec A(t) = \vec A_P(t) + \vec A_p(t)$.
We express the time-dependent Bloch function as
\begin{equation}
v_{n\vec k}(\vec r,t) = w_{n\vec k}(\vec r,t) + 
\sum_{m} C_{nm}^{\vec k}(t) w_{m \vec k}(\vec r,t),
\end{equation}
where the coefficients $C^{\vec k}_{nm}(t)$ are determined by the standard
procedure in the time-dependent perturbation theory.  We have
\begin{eqnarray}
C_{nm}^{\vec k}(t) &=& -\frac{ie}{m_ec} \int^t_{-\infty} dt' \vec P_{mn}^{\vec k}(t') \vec A_p(t')\nonumber\\
&&- \delta_{mn} \frac{ie}{m_ec}\int^t_{-\infty} dt' 
\vec k_P(t') \vec A_p(t').
\end{eqnarray}
Here we have introduced matrix elements of momentum operator,
\begin{eqnarray}
&&\vec P_{mn}^{\vec k}(t) = \int_V d\vec r w_{m\vec k}^*(\vec r,t)
\vec p w_{n\vec k}(\vec r,t) \nonumber\\
&=& (\vec p)_{nn'\vec k_P(t)}
\,\exp\left[-i\int^t_{-\infty} dt' 
(\epsilon_{n \vec k_P(t')}-\epsilon_{n' \vec k_P(t')}) \right],~~~~~
\end{eqnarray}
where 
\begin{equation}
(\vec p)_{nn'\vec k} = \int_V d\vec r u_{n\vec k}^*(\vec r) \vec{p} u_{n'\vec k}(\vec r) 
\end{equation}
is the matrix element in the static basis.

The electric current density averaged over the unit cell is given by
\begin{eqnarray}
\vec J(t) &=& -\frac{e}{m_eV}  \int_V d\vec r \nonumber\\
&&\sum_{n\vec k} 
{\rm Re} \left\{
v_{n\vec k}^* \left( \vec p + \vec k_P(t) + \frac{e}{c}\vec A_p(t) \right) 
v_{n\vec k} 
\right\}.
\end{eqnarray}
This may be decomposed into pump and probe 
contributions, $\vec J(t) = \vec J_P(t) + \vec J_p(t)$.
The two components are
\begin{equation}
\vec J_P(t) = -\frac{e}{m_eV} \sum_{n\vec k}  \int_V d\vec r
\,{\rm Re} \left\{ w_{n\vec k}^* \left( \vec p + \vec k_P(t)  \right) w_{n\vec k} 
\right\},
\end{equation}
\begin{eqnarray}
\vec J_p(t)
&=& -\frac{e^2}{m_ec} n_e \vec A_p(t) +\frac{e^2}{m_e^2 cV}\int^t_{-\infty} dt' \nonumber\\
&&\sum_{n \ne n', \vec k}{\rm Im} \left\{ \vec P_{nn'}^{\vec k}(t)  \left(\vec P_{n'n}^{\vec k}(t')\cdot \vec A_p(t')
\right)
\right\}
\label{Jtensor}
\end{eqnarray}
where $n_e$ is the average density of valence electrons. 

Equation (\ref{Jtensor}) is a linear relation between the probe 
field and the induced electric current density.  
As we described in Eq. (\ref{def_sigma}), 
we introduce a conductivity function $\sigma_{\alpha\beta}(t,t')$
that relates the probe field and the induced electric current density,
\begin{equation}
J_{\alpha}^p(t) = \sum_{\beta} \int^{\infty}_{-\infty} dt' \sigma_{\alpha\beta}(t,t')
E_{\beta}^p(t'),
\label{sigma_def}
\end{equation}
where $\alpha,\beta$ are the Cartesian indices of the current and the probe field,
and $J^p_{\alpha}$ and $E^p_{\alpha}$ indicate $\alpha$-components of probe 
electric current and probe electric field, respectively.
The conductivity function for $t>t'$ is given by
\begin{eqnarray}
&&\sigma_{\alpha\beta}(t,t') 
= \frac{e^2}{m_e} n_e \delta_{\alpha\beta} \nonumber\\
&-&\frac{e^2}{m_e^2 V}  \int^t_{t'} dt'' \sum_{n\ne n', \vec k}
 {\rm Im} \Bigg[ (p_{\alpha})_{nn' \vec k_P(t)} (p_{\beta})_{n'n \vec{k}_P(t'')}\nonumber\\
&\times&\exp\left[-i \int_{t''}^t d\tau \left\{ \epsilon_{n' \vec k_P(\tau)}
- \epsilon_{n \vec k_P(\tau)} \right\} \right] \Bigg].
\label{sigma_general}
\end{eqnarray}
This is our general expression for 
the conductivity in the presence of a strong electric field,
when the system is described by the time-dependent
Schr\"odinger equation (\ref{TDSE}).

The polarization induced by the probe field is given as the integral of the current over time.
We may introduce the linear susceptibility $\chi_{\alpha\beta}(t,t')$ by
\begin{equation}
P^p_{\alpha}(t) = \int^t_{-\infty} dt' J_p(t')
= \sum_{\beta} \int^t_{-\infty} dt' \chi_{\alpha\beta}(t,t') E^p_{\beta}(t').
\end{equation}
Therefore, the linear susceptibility and the conductivity are related by
\begin{equation}
\chi_{\alpha\beta}(t,t') = \int^t_{-\infty} dt'' \sigma_{\alpha\beta}(t'',t').
\end{equation}

\subsection{Time-resolved, frequency-dependent conductivity}

As we described in Sec. \ref{sec:Formalism},
we introduce a time-resolved, frequency-dependent conductivity by
Eq. (\ref{TD_conductivity}) using 
a probe electric field $E^p_{\beta}(t)$ which has a sharp peak at $t=T_p$,
\begin{equation}
\tilde\sigma_{\alpha\beta}(T_p,\omega) =
\frac{\tilde J^p_{\beta}(\omega)}{\tilde E^p_{\alpha}(\omega)},
\label{def_sigma2}
\end{equation}
where $\tilde J^p_{\alpha}(\omega)$ and $\tilde E^p_{\alpha}(\omega)$
are Fourier transforms of $J^p_{\alpha}(t)$ and $E^p_{\alpha}(t)$,
respectively. As a probe field, we first consider an impulsive probe 
field at $t=T_p$, $E^p_{\beta}(t) = k \delta(t-T_p)$, where $k$ specifies 
the strength of the probe pulse. With this simple choice of the probe field, 
we have the following result for the time-resolved conductivity,
\begin{equation}
\tilde\sigma^I_{\alpha\beta}(T_p,\omega) = \int ds e^{i\omega s} \sigma_{\alpha\beta}(T_p+s,T_p),
\label{TRconductivity}
\end{equation}
where the superscript $I$ of $\tilde\sigma^I_{\alpha\beta}$ indicates that
the impulsive probe field is used.
As a more general probe pulse, we consider
\begin{equation}
E_{\beta}^p(t)=f_p(t-T_p) e^{-i\omega_p (t-T_p)}
\label{general_probe}
\end{equation}
where $\omega_p$ indicates an average frequency of the probe pulse
and $f_p(t)$ is a real envelope function having a peak at $t=0$.
We assume that it is an even function, $f_p(t)=f_p(-t)$, so that the
Fourier transform, $\tilde f_p(\omega)=\int dt e^{i\omega t} f_p(t)$
is real. With this choice of the probe pulse and using Eq. (\ref{def_sigma2}), 
we have
\begin{equation}
\tilde \sigma_{\alpha\beta}(T_p,\omega)
= \frac{\int ds f_p(s) \tilde\sigma^I_{\alpha\beta}(T_p+s,\omega) e^{i(\omega-\omega_p)s}}
{\int ds f_p(s) e^{i(\omega-\omega_p)s}}.
\label{TRconductivity_general}
\end{equation}
This expression indicates that the time-resolved conductivity does not
depend much on details of the shape of the probe pulse 
if we use a short enough probe pulse $f_p$ so that 
$\tilde \sigma^I_{\alpha\beta}(T_p+s,\omega)$ does not change much
in the duration of the probe pulse.

We note that the real part of the time-resolved conductivity ${\rm Re} \tilde\sigma_{\alpha\alpha}(T_p,\omega)$ 
is related to the energy transfer from the probe electric field to electrons, as the ordinary
conductivity does. To show it, we consider the work done by the probe electric field to electrons 
which is given by
\begin{equation}
W = \int dt \vec E_p(t) \vec  J_p(t).
\end{equation}
This can be written as
\begin{eqnarray}
W &=& \sum_{\alpha} \int dt E_{\alpha}^p(t) J_{\alpha}^p(t)
\nonumber\\
&=&
\sum_{\alpha} \frac{1}{2\pi} \int d\omega \tilde E^{p*}_{\alpha}(\omega) \tilde J^p_{\alpha}(\omega)
\nonumber\\
&=&
\sum_{\alpha} \frac{1}{2\pi} \int d\omega \vert \tilde E^p_{\alpha}(\omega) \vert^2
\frac{J^p_{\alpha}(\omega)}{E^p_{\alpha}(\omega)}
\nonumber\\
&=&
\sum_{\alpha} \frac{1}{2\pi} \int d\omega \vert \tilde E^p_{\alpha}(\omega) \vert^2
\tilde \sigma_{\alpha\alpha}(T_p,\omega),
\end{eqnarray}
where we used the definition for the $\tilde\sigma_{\alpha\beta}(T_p,\omega)$ given by
Eq. (\ref{TRconductivity_general}).
Since the physical electric field, $E^p_{\alpha}(t)$, and the physical induced current,
$J^p_{\alpha}(t)$, are real quantities, we have
\begin{equation}
\tilde \sigma^*_{\alpha\alpha}(T_p,\omega) = \tilde \sigma_{\alpha\alpha}(T_p, -\omega).
\end{equation}
Using this relation, we have
\begin{eqnarray}
W &=& \sum_{\alpha} \frac{1}{2\pi} \int_{-\infty}^{\infty} d\omega \vert \tilde E^p_{\alpha}(\omega) \vert^2
\tilde \sigma_{\alpha\alpha}(T_p,\omega) 
\nonumber\\
&=& \sum_{\alpha} \frac{1}{\pi} \int_0^{\infty} d\omega \vert \tilde E^p_{\alpha}(\omega) \vert^2
{\rm Re} \tilde \sigma_{\alpha\alpha}(T_p,\omega).
\end{eqnarray}
This result clearly indicates that the energy transfer from the probe pulse to electrons
is described by the real part of the time-resolved conductivity that we defined by
Eq. (\ref{TRconductivity_general}).

 \subsection{Parabolic Two-Band Model}
 We introduce a two-band model in Eq. (\ref{sigma_general}), considering
only two orbitals in the sum, occupied valence ($v$) and unoccupied conduction
($c$) bands. The excitation energy from the valence band
to the conduction band is assumed to have a parabolic form,
\begin{equation}
\epsilon_{c \vec k} - \epsilon_{v \vec k}
\simeq \frac{k^2}{2\mu} + \epsilon_g,
\end{equation}
where $\epsilon_g$ is the band gap energy and $\mu$ is the reduced mass
of electron-hole pairs.

\subsubsection{Conductivity under a static electric field: static Franz-Keldysh effect}
For a static electric field $\vec E$, the vector potential has a linear time-dependence,
\begin{equation}
\vec A_p(t) = \vec A_p(T) - c \vec E (t-T). 
\end{equation}
The conductivity $\sigma(t,t')$ of 
Eq. (\ref{sigma_general})  is a function of $t-t'$ and the conductivity $\sigma(T_p,\omega)$ of 
Eq. (\ref{TRconductivity}) is independent of $T_p$.
After straightforward calculations, we have
\begin{eqnarray}
&&\tilde{\sigma}_{\alpha\beta}(\omega)= \frac{ie^2n_e}{m_e\omega} \delta_{\alpha\beta}
+ \frac{e^2}{m_e^2 \omega V} \int_0^{\infty} ds e^{i\omega s} \sum_{\vec k}
\nonumber\\
&& \times
\left[ (p_{\alpha})_{vc \vec k-e\vec Es/2} 
(p_{\beta})_{cv \vec k +e\vec E s/2} 
e^{-i \left\{ \left( \epsilon_k+\epsilon_g \right) s + \frac{c^2 E^2}{24\mu} s^3 \right\} } \right.
\nonumber\\
&& \left.
-  (p_{\beta})_{vc \vec k +e\vec E s/2}  (p_{\alpha})_{cv \vec k -e\vec E s/2} 
e^{i \left\{ \left( \epsilon_k+\epsilon_g \right) s + \frac{c^2 E^2}{24\mu} s^3 \right\} }
\right].
\end{eqnarray}

Below, we consider a real part of the diagonal element, ${\rm Re} \sigma_{\alpha\alpha}(\omega)$.
We further assume that $\vec k$-dependence of the matrix elements 
$(p_{\alpha})_{vc \vec k}$ may be ignored. 
Then, carrying out integration over $s$, we have
\begin{eqnarray}
{\rm Re} \tilde{\sigma}_{\alpha\alpha}(\omega)&=&
\frac{\pi e^2}{m_e^2\omega V} \vert (p_{\alpha})_{vc} \vert^2 \sum_{\vec k}
\left(\frac{8\mu}{e^2E^2} \right)^{1/3}\nonumber\\
&\times&Ai \left( (\epsilon_k+\epsilon_g-\omega) \left( \frac{8\mu}{e^2E^2} \right)^{1/3} \right),
\end{eqnarray}
where $Ai(x)$ is the Airy function.
Further carrying out $\vec k$-integration, we have
\begin{eqnarray}
&&{\rm Re} \tilde\sigma_{\alpha\alpha}(\omega)
= \frac{(2\mu)^{3/2} e^2}{2m_e^2 \omega}
\vert (p_{\alpha})_{vc} \vert^2 \sqrt{\Theta}\nonumber\\
&\times&\left\{ -\frac{\epsilon_g-\omega}{\Theta} Ai^2 \left( \frac{\epsilon_g-\omega}{\Theta} \right)
+ Ai'^2 \left( \frac{\epsilon_g-\omega}{\Theta} \right) \right\},
\label{FKE}
\end{eqnarray}
where $\Theta = (e^2 E^2 / 2\mu)^{1/3}$.
This is a well-known formula of static Franz-Keldysh effect \cite{Tharmalingam63}.

\subsubsection{Conductivity under a periodic electric field: dynamical Franz-Keldysh effect}

We next consider a case of pump electric field which is periodic in time,
$\vec A_P(t+T_{\Omega}) = \vec A_P(t)$, where $T_{\Omega}$ is the period of
the pump field and is related to the frequency $\Omega$ by $T_{\Omega}=2\pi/\Omega$. 
The conductivity $\sigma(t,t')$ has also the periodicity, 
\begin{equation}
\sigma(t,t')=\sigma(t-T_{\Omega}, t'-T_{\Omega}). 
\end{equation}
We make a Fourier expansion of $\sigma_{\alpha\beta}(t,t-s)$ which is
periodic in $t$ with the period $T_{\Omega}$,
\begin{equation}
\sigma_{\alpha\beta}(t,t-s) = \sum_{n=-\infty}^{\infty} e^{in\Omega t} \sigma_{\alpha\beta}^{(n)}(s),
\label{sigma_Fex}
\end{equation}
where $\sigma_{\alpha\beta}^{(n)}(s)$ is defined by
\begin{equation}
\sigma_{\alpha\beta}^{(n)}(s) = \frac{1}{T_{\Omega}} \int_0^{T_{\Omega}} dt
e^{-in\Omega t} \sigma_{\alpha\beta}(t,t-s).
\label{def_sigma_n}
\end{equation}
The time-resolved frequency-dependent conductivity $\tilde\sigma^I_{\alpha\beta}(T_p,\omega)$
for an impulsive probe field
defined by Eq. (\ref{TRconductivity}) may be expressed as
\begin{equation}
\tilde{\sigma}^I_{\alpha\beta}(T_p,\omega) =
\sum_n e^{in\Omega T_p} \tilde{\sigma}_{\alpha\beta}^{(n)}(\omega+n\Omega),
\label{td_sigma}
\end{equation}
where $\tilde\sigma_{\alpha\beta}^{(n)}(\omega)$ is the Fourier transform of
$\sigma_{\alpha\beta}^{(n)}(s)$.
For a general probe field of Eq. (\ref{general_probe}) with the envelope function $f_p(t)$, 
the conductivity defined by
Eq. (\ref{TRconductivity_general}) becomes
\begin{eqnarray}
\tilde \sigma_{\alpha\beta}(T_p,\omega)
&=& \frac{\int ds f(s) \tilde\sigma^I_{\alpha\beta}(T_p+s,\omega) e^{i(\omega-\omega_0)s}}
{\int ds f(s) e^{i(\omega-\omega_0)s}}
\nonumber\\
&=&
\sum_n \frac{\tilde f_p(\omega+n\Omega-\omega_0)}{\tilde f_p(\omega-\omega_0)}
e^{in\Omega T} \sigma^{(n)}_{\alpha\beta}(\omega+n\Omega),~~~~~
\label{td_sigma_general}
\end{eqnarray}
where $\tilde f_p(\omega)$ is the Fourier transform of $f_p(t)$.

We now consider a specific form of the pump vector potential,
$\vec A_P(t) = \vec A_0 \cos \Omega t$, and calculate explicit form of 
the time-resolved conductivity assuming a two-band model.
We calculate  $\tilde\sigma_{\alpha\beta}^{(n)}(\omega)$ in two steps.
We first calculate the Fourier transform of $\sigma_{\alpha\beta}(t,t-s)$ with respect to $s$
which we denote as $\hat\sigma_{\alpha\beta}(t,\omega)$.
In the parabolic two-band model, it is given as follows.
\begin{eqnarray}
&&\hat{\sigma}_{\alpha\beta}(t,\omega)
= \int_{-\infty}^{\infty} ds e^{i\omega s} \sigma_{\alpha\beta}(t,t-s)
\nonumber\\
&=&
\frac{i e^2}{m_e \omega} n_e \delta_{\alpha\beta}
+\frac{e^2}{m_e^2 \omega V} \int^{\infty}_0 ds e^{i\omega s} \sum_{\vec k}
\nonumber\\
&& \left[ (p_{\alpha})_{vc} (p_{\beta})_{cv}
e^{-i \int_0^s dy \left\{ \frac{1}{2\mu} \left( \vec k + \frac{e}{c}\vec A(t-y) \right)^2 + \epsilon_g \right\} }
\right.
\nonumber\\
&& - \left. (p_{\beta})_{vc} (p_{\alpha})_{cv}
e^{i \int_0^s dy \left\{ \frac{1}{2\mu} \left( \vec k + \frac{e}{c}\vec A(t-y) \right)^2 + \epsilon_g \right\} } \right].
\label{sigma_temp}
\end{eqnarray}
The integral in the exponential is calculated as
\begin{eqnarray}
&&\int_0^s dy \left\{ \frac{1}{2\mu} \left( \vec k + \frac{e}{c} \vec A(t-y) \right)^2 
+ \epsilon_g \right\} \nonumber\\
&=&
\left( \epsilon_k + \epsilon_g + U_p \right) s
-\theta_1 \sin \Omega(t-s) + \theta_1 \sin \Omega t \nonumber\\
&&-\theta_2 \sin 2\Omega (t-s) + \theta_2 \sin 2\Omega t,
\end{eqnarray}
where we introduced
\begin{equation}
U_p = \frac{e^2 A_0^2}{4\mu c^2},
\end{equation}
\begin{equation}
\theta_1 = \frac{e \vec k \cdot \vec A_0}{\mu c \Omega},
\end{equation}
\begin{equation}
\theta_2 = \frac{e^2 A_0^2}{8\mu c^2 \Omega}.
\end{equation}

Using a relation involving Bessel function $J_n(x)$,
\begin{equation}
e^{ia\sin\theta} = \sum_n J_n(a) e^{in\theta},
\end{equation}
where $n$ runs for whole integers, we may express
\begin{eqnarray}
e^{i\theta_1 \sin \Omega t + i\theta_2 \sin 2\Omega t}
&=& \sum_{l' m} J_{l'}(\theta_1) J_m(\theta_2) e^{i(l'+2m) \Omega t}
\nonumber\\
&=& \sum_l J_l(\theta_1,\theta_2) e^{il\Omega t},
\end{eqnarray}
where we defined the generalized Bessen function by
\begin{equation}
J_l(\theta_1,\theta_2) = \sum_m J_{l-2m}(\theta_1) J_m(\theta_2).
\label{gbessel}
\end{equation}
We may express the integral in Eq. (\ref{sigma_temp}) as
\begin{eqnarray}
&& \int_0^{\infty} ds e^{i\omega s} 
\exp\Big[-i \big[ (\epsilon_k+\epsilon_g+U_p)s -\theta_1 \sin \Omega(t-s) \nonumber\\
&&+ \theta_1 \sin \Omega t-\theta_s \sin 2\Omega (t-s) + \theta_2 2\Omega t \big]\Big]
\nonumber\\
&=&
\sum_{l_1 l_2} 
\frac{i J_{l_1}(\theta_1,\theta_2) J_{l_2}(\theta_1,\theta_2)}
{\omega-(\epsilon_k + \epsilon_g + U_p + l_1\Omega )} e^{i(l_1-l_2)\Omega t}.
\end{eqnarray}
Equation (\ref{sigma_temp}) is now written as
\begin{eqnarray}
&&\hat{\sigma}_{\alpha\beta}(t,\omega)
=
\frac{i e^2}{m_e \omega} n_e \delta_{\alpha\beta}
+\frac{i e^2}{m_e^2 \omega V} \sum_{\vec k l_1 l_2} J_{l_1}(\theta_1,\theta_2) J_{l_2}(\theta_1,\theta_2)
\nonumber\\
&& \times\Bigg[ (p_{\alpha})_{vc} (p_{\beta})_{cv}
\frac{e^{i(l_1-l_2) \Omega t}}{\omega-(\epsilon_k+\epsilon_g+U_p+l_1\Omega)} \nonumber\\
&&-(p_{\beta})_{vc} (p_{\alpha})_{cv}
\frac{e^{-i(l_1-l_2) \Omega t}}{\omega+(\epsilon_k+\epsilon_g+U_p+l_1\Omega)}
 \Bigg].
\end{eqnarray}

We next calculate $\tilde{\sigma}_{\alpha\beta}^{(n)}(\omega)$ as follows.
\begin{eqnarray}
&&\tilde{\sigma}_{\alpha\beta}^{(n)}(\omega) 
=
\frac{1}{T_{\Omega}} \int_0^{T_{\Omega}} dt e^{-in\Omega t} \hat{\sigma}_{\alpha\beta}(t,\omega)
\nonumber\\
&=&
\frac{ie^2}{m_e\omega} n_e \delta_{\alpha\beta} \delta_{n0}
+ \frac{ie^2}{m_e^2 \omega V} \sum_{\vec k l} 
\nonumber\\
&& \Bigg[ J_l(\theta_1,\theta_2) J_{l-n}(\theta_1,\theta_2)
\frac{(p_{\alpha})_{vc}(p_{\beta})_{cv}}{\omega-(\epsilon_k+\epsilon_g+U_p+l\Omega)} 
\nonumber\\
 &-& J_l(\theta_1,\theta_2) J_{l+n}(\theta_1,\theta_2)
\frac{(p_{\beta})_{vc}(p_{\alpha})_{cv}}{\omega+(\epsilon_k+\epsilon_g+U_p+l\Omega)}
\Bigg].~~~
\label{sigma_n_cos}
\end{eqnarray}
We put the above expression into Eq. (\ref{td_sigma_general}),
\begin{eqnarray}
\tilde{\sigma}_{\alpha\beta}(T_p,\omega)
&=&
\sum_n \frac{\tilde f_p(\omega+n\Omega-\omega_0)}{\tilde f_p(\omega-\omega_0)} e^{in\Omega T_p}
\tilde{\sigma}_{\alpha\beta}^{(n)}(\omega+n\Omega)
\nonumber\\
&=&
\frac{ie^2}{m_e\omega} n_e \delta_{\alpha\beta} \nonumber\\
&+& \sum_{\vec k l n}\frac{ie^2}{m_e^2 (\omega+n\Omega) V} 
\frac{\tilde f_p(\omega+n\Omega-\omega_0)}{\tilde f_p(\omega-\omega_0)} e^{in\Omega T_p}
\nonumber\\
& \times& \Bigg[ 
\frac{(p_{\alpha})_{vc}(p_{\beta})_{cv}J_l(\theta_1,\theta_2) J_{l-n}(\theta_1,\theta_2)}
{\omega+n\Omega-(\epsilon_k+\epsilon_g+U_p+l\Omega)} 
\nonumber\\
&&- \frac{(p_{\beta})_{vc}(p_{\alpha})_{cv}J_l(\theta_1,\theta_2) J_{l+n}(\theta_1,\theta_2)}
{\omega+n\Omega+(\epsilon_k+\epsilon_g+U_p+l\Omega)}
\Bigg].
\label{final_conductivity}
\end{eqnarray}

We carry out $\vec k$ integration in terms of $\epsilon_k = k^2/2\mu$
and $\cos\theta_k$. After integration over $\cos\theta_k$, only even $n$ terms contribute.
We introduce the following quantity,
\begin{equation}
\xi_{ln}(k)
=\int_{-1}^{1} d(\cos\theta_k) 
J_l(\theta_1,\theta_2) J_{l-n}(\theta_1,\theta_2),
\end{equation}
and we denote $n=2m$ below.
For the diagonal term $\alpha=\beta$, we have
\begin{eqnarray}
&&\tilde{\sigma}_{\alpha\alpha}(T_p,\omega)
=
\frac{ie^2}{m_e\omega} n_e 
\nonumber\\
&+& \sum_{l m}\frac{ie^2 \mu^{3/2}\vert (p_{\alpha})_{vc} \vert^2}{\sqrt{2} \pi^2 m_e^2(\omega+2m\Omega)} 
\frac{\tilde f_p(\omega+2m\Omega-\omega_0)}{\tilde f_p(\omega-\omega_0)} e^{i2m\Omega T_p}
\nonumber\\
&\times&
\int_0^{\infty} \sqrt{\epsilon_k} d\epsilon_k 
\Bigg[ \frac{\xi_{l,2m(k)}}
{\omega+2m\Omega-(\epsilon_k+\epsilon_g+U_p) - l\Omega}
\nonumber\\
&&-\frac{\xi_{l,-2m}(k)}
{\omega+2m\Omega+(\epsilon_k+\epsilon_g+U_p)+l\Omega}
\Bigg].
\end{eqnarray}

As the final result of this subsection, we obtain the
following expression for the real part of the conductivity,
\begin{eqnarray}
&&{\rm Re} \tilde{\sigma}_{\alpha\alpha}(T_p,\omega)
= 
 \sum_{m=-\infty}^{\infty}\frac{e^2 \mu^{3/2}\vert (p_{\alpha})_{vc} \vert^2}{\sqrt{2} \pi^2 m_e^2 (\omega+2m\Omega)} 
\nonumber\\
&&\times\frac{\tilde f_p(\omega + 2m\Omega-\omega_0)}{\tilde f_p(\omega-\omega_0)}\big[
 C_m(\omega)  \cos 2m\Omega T_p \nonumber\\
 &&+ S_m(\omega) \sin 2m\Omega T_p\big].
\label{sigma_CS}
\end{eqnarray}
Coefficients $C_m(\omega)$ are given by
\begin{eqnarray}
&&C_m(\omega)=
\pi \int_0^{\infty} \sqrt{\epsilon_k} d\epsilon_k \nonumber\\
&\times& \sum_{l} \left\{ \xi_{l,2m}(k)
\delta(\omega+2m\Omega-(\epsilon_k+\epsilon_g+U_p +l \Omega)) \right.
\nonumber\\
&& \left.
-\xi_{l,-2m}(k) \delta(\omega+2m\Omega+(\epsilon_k+\epsilon_g+U_p+l\Omega)) \right\}
\nonumber\\
&=&
\sum_l \pi \left[ 
\sqrt{\epsilon_k^+} \xi_{l,2m}(k^+)
-
\sqrt{\epsilon_k^-} \xi_{l,-2m}(k^-) \right],
\label{Cm}
\end{eqnarray}
where $k$ and $k^{\pm}$ are related to $\epsilon_k$ and
$\epsilon_k^{\pm}$ by $k=\sqrt{2\mu \epsilon_k}$.
$\epsilon_k^{\pm}$ are defined by
\begin{equation}
\epsilon_k^{\pm} =
\pm (\omega + 2m\Omega) - (\epsilon_g + U_p +l \Omega).
\end{equation}
Coefficients $S_m(\omega)$ are given by
\begin{eqnarray}
&&S_m(\omega)
=
-\int_0^{\infty} \sqrt{\epsilon_k} d\epsilon_k \nonumber\\
&\times&\sum_l \Bigg[
\frac{\xi_{l,2m}(k)}{\omega+2m\Omega-(\epsilon_k+\epsilon_g+U_p+l\Omega)} \nonumber\\ 
&&+\frac{\xi_{l,-2m}(k)}{\omega-2m\Omega+(\epsilon_k+\epsilon_g+U_p+l\Omega)} \Bigg].
\label{Sm}
\end{eqnarray}
We note that the terms with $m=\pm 1$ are responsible for the phase shift 
seen in Fig. \ref{fig:TD_Sigma}.

The above expression of the time-resolved conductivity depends on 
the choice of the envelope function $f_p(t)$ through
$\tilde{f}_p(\omega)$.
Modulations with a long period are influenced by 
$\tilde{f}_p(\omega)$ at small $\omega$, typically 
$\omega \simeq\pm 2\Omega$.
This indicates that the conductivity does not 
depend much on the detail of the probe pulse, if we 
employ a probe pulse which is much shorter than 
the period $T$ of the pump. In the TDDFT calculation 
shown in Fig. 2, we used a short enough probe pulse
with $\tilde f_p(\pm 2\Omega)/f_p(0)=0.96$ close to unity.

 
  \subsection{Numerical results}
  
 \begin{figure} 
\includegraphics[width=90mm]{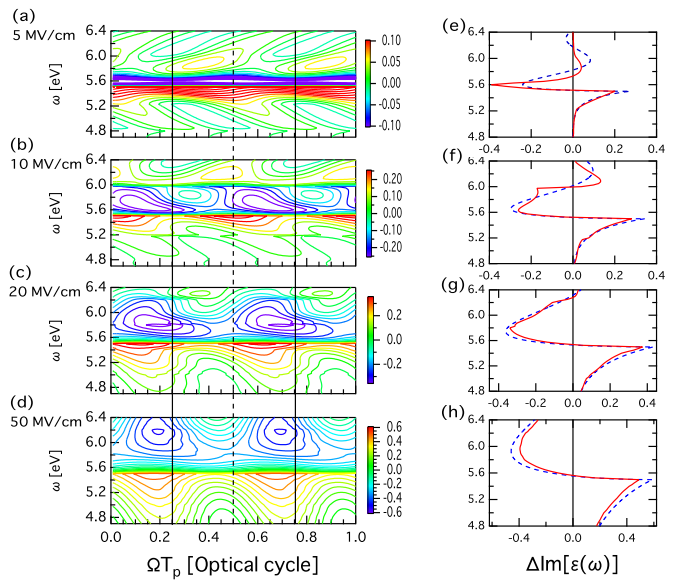} 
\caption{\label{fig:Houston} Contour plots and time averages of 
$\Delta {\rm Im}[\varepsilon(\omega,T_p)]$ in the two-band model 
for the pump field intensities of (a,e)$E_0=5$, (b,f) 10,  (c,g) 20, and (d,h) 50 MV/cm.   
In the left panels (a)-(d), the horizon axis is the phase defined by $\Omega T_p$. 
The vertical solid (dashed) lines indicate the position of maximum (minimum)  
of the electric field. In the right panels (e)-(h), red-solid lines is the time average 
of the modulation. Blue-dotted lines show result of static FKE.} 
\end{figure}

We numerically calculate the change of the dielectric function  
$ \Delta{\rm Im}[\varepsilon(\omega,T_p)]$ 
using Eqs. (\ref{sigma_CS}) - (\ref{Sm}). 
In the calculations shown below, we simply put 
$\tilde f(\omega)=1$. This is equivalent to using
an impulsive field for the probe, $E_p(t) \propto
\delta (t-T_p)$.

Figure \ref{fig:Houston} (a)-(d) show  
the contour plot, where the horizontal axis is the phase defined by $\Omega T_p$. 
In Fig. \ref{fig:Houston} (e)-(h), modulations averaged 
over time is shown by red-solid lines. Blue-dashed lines show results of 
static FKE: the time-averaged modulation using static formula
, Eq.~(\ref{FKE}) 
where the parameter $\theta$ is evaluated using time-dependent 
electric field, $E(t)$. 
The parameters in the two-band model are set to reproduce responses 
in the TDDFT calculations: 
The frequency of pump field is set to $\Omega=0.4$ eV
, the reduced mass is set to $\mu=0.25m_e$, 
and 
the dipole matrix element  
is set to $\vert p_{cv} \vert^2 = 0.928$ in atomic unit.

We first look at time-averaged modulations shown in panels (e)-(h).  
In all panels, we find appearances of absorption below the 
band gap and decrease above the band gap.  
We can also find structures of at $\omega=\epsilon_g \pm \Omega$,
most clearly in (b) and (f), which are cause by side band contributions
with different $l$ values in Eqs. (\ref{Cm}) and (\ref{Sm}).
They show a similar behavior to Fig.~{\ref{fig:TD_Sigma} (g)-(k). 
We also find that the time-averaged modulation is quite close to 
the estimation by the static FKE, even at the smallest intensity 
where the $\gamma$ value is much less than unity. 
We repeat a similar calculation changing the frequency of
the pump field while the intensity is fixed at 50 MV/cm, and have found
that the static FKE describes reasonably the time-averaged modulations
below the band gap.
 
We next look at the time dependence of the modulation. 
At the strong pump electric field (d),  we find a large 
modulation when the magnitude of the pump electric field is large. 
We find a phase shift as the pump electric field decreases. 
The maximum of the modulation moves forward in time. 
From these observations, we can say that all the features  
seen in the first-principles TDDFT calculations shown in  
Fig. \ref{fig:TD_Sigma} are reproduced by the analytic  
formula of Eqs. (\ref{sigma_CS}) - (\ref{Sm}) of the simple two-band model.

\section{Summary }
We have developed a theoretical pump-probe formalism 
to investigate the dynamical Franz-Keldysh effect in time-domain, 
in femtosecond time resolution much shorter than the optical 
cycle of the applied pump field. 
Both numerical simulations based on real-time time-dependent 
density functional theory and an analytic approach in the two-band 
model reveal the same behavior for the modulations of dielectric  
properties. We find the time-averaged behavior in the DFKE can be 
well described by the static FKE. The most remarkable feature of 
the DFKE is that there appears a phase shift between the modulation of 
the dielectric function and the applied pump field which becomes 
significant as the magnitude of the electric field decreases.


\section*{acknowledgments}
The authors thank G.F. Bertsch for discussions and for carefully reading manuscript.
This work is supported by a Grant-in-Aid for Scientific 
Research (No. 21740303 and No. 15H03674). 
Numerical calculations were performed on the supercomputer PRIMARGY BX900 at 
the Japan Atomic Energy Agency (JAEA) and   
the K computer at the RIKEN Advanced Institute for Computational Science 
(proposal number hp120065).


\begin{thebibliography}{99} 
\bibitem{atto01} M. Hentschel, R. Klenberger, Ch. Spielmann, G.A. Reider, N. Milosevic, T. Brabec, U. Heinzmann,
M. Drescher, and F. Krausz, Nature {\bf 414}, 509 (2001). 
\bibitem{HDBT11} H. Hirori, A. Doi, F. Blanchard, and K. Tanaka, Appli. Phys. Lett. {\bf 98}, 091106 (2011). 
\bibitem{Chin01} A.H. Chin, O.G. Calderon, and J. Kono, Phys.Rev. Lett. {\bf 86} , 3292 (2001). 
\bibitem{Hirori11} H.Hirori, K. Shinokita, M. Shirai, S. Tani, Y. Kadoya, and K. Tanaka, Nat. Comm. {\bf 2}, 594 (2011).
\bibitem{Krausz13} A. Schiffrin, T. Paasch-Colberg, N. Karpowicz, V. Apalkov, D. Gerster,
S. M\"uhlbrandt, M. Korbman, J. Reichert, M. Schultze, S. Holzner, J.V. Barth, R. Kienberger, R. Ernstorfer, V.S. Yakovlev,
M.I. Stockman, and F. Krausz, Nature {\bf 493}, 70 (2013).
\bibitem{Schultze13} M. Schultze, E.M. Botschafter, A. Sommer, S. Holzner, W. Schweinberger, M. Fless, M. Hofstetter,
R. Kienberger, V. Apalkov, V.S. Yakovlev, M.I. Stockman, F. Krausz, Nature {\bf 493}, 75 (2013).
\bibitem{Novelli13} F. Nobelli, D. Fausti, F. Giusti, F. Parmigiani, and M. Hoffmann, Scientific Reports {\bf 3} 1227 (2013). 
\bibitem{Schultze14} M. Schultze, K. Ramasesha, C.D. Pemmaraju, S.A. Sato, D. Whitmore, A. Gandman, J.S. Prell,
L.J. Borja, D. Prendergast, K. Yabana, D.M. Neumark, S.R. Leone, Science {\bf 346}, 1348 (2014).
\bibitem{Franz58} W.Franz, Z. Naturforsch. Teil A {\bf 13}, 484 (1958). 
\bibitem{Keldysh58} L. V. Keldysh, Sov. Phys. JETP {\bf 34}, 788 (1958). 
\bibitem{Tharmalingam63} K. Tharmalingam, Phys. Rev. {\bf130}, 2204 (1963). 
\bibitem{Seraphin65} B. O. Seraphin and R. B. Hess, Phys. Rev. Lett. {\bf 14}, 138 (1965). 
\bibitem{Nahory68} R. E. Nahory and J. L. Shay, Phys. Rev. Lett. {\bf 21}, 1569 (1968). 
\bibitem{Shen95} H.Shen, M. Dutta, J. Appl. Phys. {\bf 78}, 2151 (1995).
\bibitem{Sipe10} J.K. Wahlstrand, J.E. Sipe, Phys. Rev. B{\bf 82}, 075206 (2010).
\bibitem{Sipe15} F. Duque-Gomez, J.E. Sipe, J. Phys. Chem. Solids {\bf 76}, 138 (2015).
\bibitem{Yacoby68} Y. Yacoby, Phys. Rev. {\bf 169}, 610 (1968). 
\bibitem{Jauho96} A. P. Jauho and K. Johnsen, Phys.Rev. Lett. {\bf76}, 4576 (1996). 
\bibitem{Nordstorm98} K. B. Nordstrom, K. Johnsen, S. J. Allen, A. P. Jauho, B. Birnir, J. Kono, T. Noda, H. Akiyama, and H. Sakaki, 
Phys. Rev. Lett. {\bf 81}, 457 (1998).
\bibitem{Ajit04} Ajit Srivastava, Rahul Srivastava, Jigang Wang, and Junichiro Kono, Phys.Rev.Lett. {\bf 93}, 157401 (2004). 
\bibitem{Mizumoto06} Y. Mizumoto, Y. Kayanuma, A. Srivastava, J. Kono, and A. H. Chin, Phys. Rev. B {\bf 74}, 045216 (2006). 
\bibitem{Shambhu11} S. Ghimire, A.D. DiChiara, E. Sistrunk, U.B. Szafruga, P. Agostini, L. F. DiMauro, and D. A. Reis, Phys. Rev. Lett. {\bf 107}, 167407 (2011). 
\bibitem{Chin00} A. H. Chin, J. M. Bakker, and J. Kono, Phys. Rev. Lett. {\bf 85}, 3293 (2000). 
\bibitem{Srivastava04} Ajit Srivastava, Rahul Srivastava, JigangWang, and Junichiro Kono, Phys. Rev. Lett. {\bf 93}, 157401 (2004).  
\bibitem{Runge84} E. Runge and E. K. U. Gross,  Phys.Rev. Lett. {\bf 52}, 997 (1984). 
\bibitem{Bertsch00} G.F. Bertsch, J.-I. Iwata, A. Rubio, and K. Yabana, Phys. Rev. B {\bf62} , 7998 (2000). 
\bibitem{TM91} N. Troullier and J.L. Martins, Phys. Rev. {\bf B43}, 1993 (1991). 
\bibitem{Otobe08}  
T. Otobe, M. Yamagiwa, J. -I. Iwata, K. Yabana, T. Nakatsukasa, and G. F. Bertsch,  
Phys. Rev. B{\bf77}, 165104 (2008). 
\bibitem{Otobe09}T. Otobe, K. Yabana, J.-I. Iwata, J. Phys.: Condens. Matter. {\bf 21}, 064224 (2009).
\bibitem{Shino10-2} Y. Shinohara, K. Yabana, Y. Kawashita, J.-I. Iwata, T. Otobe , and George F. Bertsch,  Phys. Rev. B {\bf 82}, 155110 (2010). 
\bibitem{Shino12} Y. Shinohara, S. A. Sato, K. Yabana, J.-I. Iwata, and T. Otobe, 
	        J. Chem. Phys. {\bf 137} 22A527 (2012). 
\bibitem{Otobe12} T. Otobe, J. Appli. Phys. {\bf 111}, 093112 (2012). 
\bibitem{Sato14} S.A. Sato, K. Yabana, Y. Shinohara, T. Otobe, G.F. Bertsch, 
Phys. Rev. {\bf B89}, 064304 (2014). 
\bibitem{Wachter14} G. Wachter, C. Lemell, J. Burgdorfer, S.A. Sato, X.M. Tong, K. Yabana, 
Phys. Rev. Lett. {\bf 113}, 087401 (2014). 
\bibitem{TM91} N. Troullier and J.L. Martins, Phys. Rev. {\bf B43}, 1993 (1991). 
\bibitem{Kleinman82} 
L. Kleinman and D. M. Bylander, Phys. Rev. Lett.  {\bf 48}, 1425 (1982). 
\bibitem{PZ81} J. P. Perdew and A. Zunger, Phys. Rev. B  {\bf 23}, 5048 (1981). 
\bibitem{Yabana96} K. Yabana and G.F. Bertsch, Phys. Rev. {\bf B54}, 4484 (1996). 
\bibitem{yabana12} K. Yabana, T. Sugiyama, Y. Shinohara, T. Otobe, and G. F. Bertsch, Phys. Rev. B  {\bf 85}, 045134 (2012). 
\bibitem{Vignale96} G. Vignale and W. Kohn, Phys. Rev. Lett. {\bf 77}, 2037 (1996). 
\bibitem{yabana06} K. Yabana, T. Nakatsukasa, J.-I. Iwata, and G. F. Bertsch, Phys. Stat. Solidi (b) {\bf 243}, 1121 (2006)
\bibitem{Houston} W. V.  Houston, Phys. Rev. {\bf 51}, 184 (1940).  

\end{thebibliography}
\end{document}